# Valley-contrasting orbital angular momentum in photonic valley crystals

Xiao-Dong Chen, Min Chen, and Jian-Wen Dong*

*School of Physics & State Key Laboratory of Optoelectronic Materials and Technologies, Sun Yat-Sen University, Guangzhou 510275, China.*

*Corresponding author: dongjwen@mail.sysu.edu.cn

**Valley, as a degree of freedom, has been exploited to realize valley-selective Hall transport and circular dichroism in two-dimensional layered materials. On the other hand, orbital angular momentum of light with helical phase distribution has attracted great attention for its unprecedented opportunity to optical communicagtions, atom trapping, and even nontrivial topology engineering. Here, we reveal valley-contrasting orbital angular momentum in all-dielectric photonic valley crystals. Selective excitation of valley chiral bulk states is realized by sources carrying orbital angular momentum with proper chirality. Valley dependent edge states, predictable by nonzero valley Chern number, enable to suppress the inter-valley scattering along zigzag boundary, leading to broadband robust transmission in Z-shape bend without corner morphological optimization. Our work may open up a new door towards the discovery of novel quantum states and the manipulation of spin-orbit interaction of light in nanophotonics.**

Valley – the energy extrema of band structure at momentum space, is a ubiquitous discrete degree of freedom in periodic systems. It has been employed to achieve a number of intriguing phenomena such as valley-selective Hall transport and circular dichroism in two-dimensional monolayer electronic materials[1-10], in which valley plays an important role in valleytronics[11] in a similar to spin in spintronics. On the other hand, valley in photonics is seldom to connect with spin-orbit interaction of light, in particular for orbital angular momentum (OAM) of light[12,13]. Utilizing duality-symmetry-protected metamaterials and spatially dependent structured metasurface seem to be necessary for spin-orbital interaction of light[14-17]. Such complexity will impede spin-orbit interaction of light in full wave spectrum, limiting to achieve nontrivial light manipulations not only in various applications fields of OAM communications, 3D particle trapping, and unidirectional transport, but also in fundamental study such as quantum mechanics simulations[18-26].

In this paper, we reveal valley-contrasting orbital angular momentum in photonic valley crystals made of all-dielectric medium, such as silicon. Using valley as a binary degree of freedom, the selective excitation of valley chiral bulk states is realized by sources carrying OAM with proper chirality. Nontrivial topological phase transition is also illustrated. Valley dependent edge states determined by nonzero valley Chern number, enable to suppress inter-valley scattering along zigzag boundary. It leads to valley-protected broadband (with a bandwidth of 10%) robust transport in a Z-shape bend photonic crystal waveguide without corner morphological optimization.

Consider a two-dimensional photonic valley crystal, which is a composite of two interlaced triangular lattices with same lattice constants (Fig. 1a). The unit-cell consists of two silicon rods embedded in air background, e.g., A and B rods in dash-rhomboid or dash-hexagon. The A rod has a diameter of $d_A = 0.5a$ while the B rod has a diameter of $d_B = 0.38a$, where $a$ is the lattice constant. The relative permittivity of each rod is $\varepsilon = 11.7$ corresponding to that of silicon at the wavelength of 1550nm. Figure 1b shows two lowest bulk bands of transverse-magnetic states with nonzero out-of-plane component of electric fields ($E_z$) and nonzero in-plane component of magnetic fields ($\vec{H}_\parallel$). Clearly, a complete band gap appears due to inversion asymmetry ($d_A \neq d_B$). In addition, it shows two pairs of valley states with the frequency of 0.244 and 0.272 $c/a$ locating at two inequivalent but time-reversal Brillouin zone corners (K' and K).

An interesting issue is that such photonic valley states can show exotic chirality such that each valley state has an intrinsic circular polarized OAM, which is verified by the phase distribution of $E_z$, i.e., arg($E_z$). In Fig. 1c, the $E_z$ phases of K' valley state decrease counterclockwise around the high symmetric central point of the unit-cell. It indicates that the $E_z$ field will rotate counterclockwise as time evolves (see the animation in Supplementary Information). In this way, the $E_z$ field has a left-handed circular polarized (LCP) OAM, marked by 3D schematic in the left of Fig. 1c. In contrast, as clearly illustrated in Fig. 1d, the $E_z$ phases of K valley state is reversed to clockwise decreasing, showing that the $E_z$ field has a right-handed circular polarized (RCP) OAM.

Inferring from the phase distributions, valley-contrasting OAM is found between two valley states at the same band but different valleys. Such valley dependence of OAM chirality can be used to achieve selective excitation of valley chiral states. It should be noted that the circular polarized OAM can also be found around the center of either A or B rod. In this paper, we only focus on the discussion of OAM chirality around the center of the unit-cell. Furthermore, as correlated by Maxwell equations, such $E_z$ field with circular polarized OAM always leads to $\bar{\mathbf{H}}_\parallel$ with circular polarized spin angular momentum at origin, and see Supplementary Information SA for more detailed discussion.

The valley chiral states can be selectively excited by external sources with proper chirality. As depicted in Fig. 2a, a circular $E_z$ line source (yellow marker, inset) with same magnitudes but phases carrying chiral OAM is excited around the center of a hexagonal photonic valley crystal with six zigzag boundaries. Note that at the boundaries, the high order Bragg-diffractions of valley chiral states are suppressed as the hexagonal photonic valley crystal is surrounded by homogeneous medium with a refractive index of 1.5, e.g., $SiO_2$[27]. The operating frequency is set to be 0.244 $c/a$. It is known that the valley chiral state will be excited only when its chirality matches the chirality of $E_z$ source. So when the phases of $E_z$ source decrease counterclockwise, i.e., LCP chirality (left panel in Fig. 2b), it couples to the K' valley state. At each boundary, the excited K' valley state is partly reflected back into the photonic valley crystal and partly refracted into the surrounding $SiO_2$ medium according to the conservation of the momentum parallel to the boundary (Fig. 2c). Particularly, the refracted light-beams are interfered and enhanced at the left-most, upper-right and lower-right corners of the sample. In addition, the phases of $E_z$ field inside the photonic valley crystal show LCP chirality, confirming that K' valley state is excited (Fig. 2d). When the phases of $E_z$ source have opposite chirality, i.e., RCP chirality as shown in the right panel of Fig. 2b, it will couple to the K valley states and the output directions of the refracted light-beams switch to enhance at the right-most, upper-left and lower-left corners of the sample (Fig. 2e). The excited $E_z$ phases will show RCP chirality, confirming that K valley state is excited (Fig. 2f). Such distinct differences of the refracted light-beam directions and the phase distributions are experimentally detectable, and in turn prove the selective excitation of valley chiral states. Note that the valley chiral states can be also selectively excited by an in-plane point-source with chiral spin angular momentum and the detailed results are presented in Supplementary Information SB.

Nontrivial topology in photonic crystals has been widely studied in gyroelectric, magnetic and bianisotropic media[28-30]. In the following, we will show that the valley chiral states give birth to unique topological charge distributions. It leads to a new nontrivial topological phase characterized by nonzero valley Chern number. In order to have a deep understanding on the topological role of valley chiral states, we construct a minimal band model of bulk dispersions. Through the **k·p** approximation, we can resort to the photonic effective Hamiltonian:

$$\hat{H} = v_D(\hat{\sigma}_x \hat{\tau}_z \delta k_x + \hat{\sigma}_y \delta k_y) + \lambda^P_{\varepsilon_z} \hat{\sigma}_z \qquad (1)$$

where $\delta \vec{k}$ is measured from the valley center K' or K point. $\hat{\sigma}_i$ and $\hat{\tau}_i$ are the Pauli matrices acting on sub-lattice and valley spaces, respectively [see detailed derivation in Supplementary Information SC]. The first two terms in the effective

Hamiltonian imply the gapless conical dispersions with Dirac velocity $v_D$ when the inversion symmetry is preserved. In contrast, $\lambda^P_{\varepsilon_z}\hat{\sigma}_z$ opens a frequency band gap with a bandwidth of $2\lambda^P_{\varepsilon_z}$ at valley centers due to the inversion asymmetry. Specifically, $\lambda^P_{\varepsilon_z} \propto [\int_B \varepsilon_z ds - \int_A \varepsilon_z ds]$ where $\int \varepsilon_z ds$ denotes the integration of $\varepsilon_z$ at A or B rod domain. Take the photonic valley crystal with $d_A = 0.5a$ and $d_B = 0.38a$ in Fig. 1 as an example, $\int_B \varepsilon_z ds < \int_A \varepsilon_z ds$ leads to $\lambda^P_{\varepsilon_z} < 0$ and thus a complete band gap is clearly presented in Fig. 1b. Moreover, the effective Hamiltonian of Equation (1) implies that the Berry curvature has a distribution sharply centered at the two valleys[1]: $\Omega(\delta \vec{k}) = \tau_z \cdot v_D^2 \lambda^P_{\varepsilon_z} / 2[(\lambda^P_{\varepsilon_z})^2 + (v_D \cdot \delta \vec{k})^2]^{3/2}$. Each valley carries a valley dependent nonzero topological charge with opposite signs $C_{\tau_z} = \tau_z \, \text{sgn}(\lambda^P_{\varepsilon_z})/2$ because the valley chiral states at the same band have opposite chirality. Then the topological charges at K' and K valleys are $C_{K'} = +1/2$ and $C_K = -1/2$, schematically denoted by white '+' and '-' symbols in the right panel of Fig. 3a. As a result, the inversion asymmetric photonic valley crystal does acquire the nonzero topological invariant -- valley Chern number $C_V = (C_K - C_{K'}) = -1 \neq 0$, although the total Chern number $C = (C_K + C_{K'})$ is zero in such time-reversal invariant system.

Indicating by the topological charge distribution function $C_{\tau_z}$, the topological phase transition accompanied with the change of valley Chern number is expected when the sign of $\lambda^P_{\varepsilon_z}$ is flipped. Here, we change the diameter of the two silicon rods to study different photonic valley crystals with inversion asymmetry amplitude denoted by $\delta d = (d_A - d_B)/2$. Figure 3b plots the frequency spectra of valley chiral states as a function of $\delta d$, while keeping the average diameter $(d_A + d_B)/2$ unchanged.

With the decrease of δ*d*, the radius of A rod becomes smaller while that of B rod goes to larger (see in bottom insets of Fig. 3b). In addition, the valley frequencies of LCP state at K' valley and RCP state at K valley will boost. On the other hand, the frequencies of the other two valley chiral states, i.e., RCP state at K' valley and LCP state at K valley, will drop with the decrease of δ*d*. The state exchange between valley states with opposite chirality at the same valley will occur at the critical point of δ*d* = 0*a*, causing the sign flipping of $\lambda_{\varepsilon_2}^P$ and the reversing of topological charge distributions. Left panel of Fig. 3a illustrates that the topological charge is +1/2 at K valley (red) and -1/2 at K' valley (blue) for photonic valley crystal with δ*d* < 0*a*, resulting in a nonzero positive valley Chern number of $C_v$ = +1, which are totally different to those of δ*d* > 0*a*. Hence, the topological invariant of valley Chern number changes and the topological phase transition is well demonstrated in photonic valley crystal.

The nonzero valley Chern number shows the intrinsic nontrivial topology in photonic valley crystal, implying to achieve robust edge states at the boundary of two topologically distinct structures. Figure 4a shows the schematic of zigzag edge constructed by photonic valley crystal with δ*d* = -0.06*a* at the bottom (orange) and another one with δ*d* = 0.06*a* (green) at the top. It is found in Fig. 3, that the orange one has a topological charge distribution of $C_{K'}$ = -1/2 and $C_K$ = +1/2, while the green one has a reversal distribution, leading that the topological charge differences at K' and K valleys are +1 and -1 when crossing the zigzag edge. According to the bulk-edge correspondence[31], there will be one edge state with positive velocity at K' valley and the other with negative velocity at K valley. Such valley dependent edge states are illustrated in Fig. 4b. Note that the dispersion of the edge states can be controlled by interchanging the relative positions of two crystals.

We would like to emphasize that the valley dependent edge states *do not* have the same origin as those in photonic quantum Hall effect by breaking time-reversal symmetry[28,29], as the Chern number in photonic valley crystal is definitely zero. This is also why the edge dispersions in Fig. 4b do not connect the valence and conduction bulk bands and they are gapped. Although the valley dependent edge states are gapped, they can be employed for constructing waveguide with broadband robust transport against sharp corners, as the inter-valley scattering is suppressed due to the vanishing field overlapping between two valley states[32]. This valley-assisted way for robust transport is somehow superior to the means of topological photonic metamaterials and metacrystals, due to all-dielectric constituents. As a representative example, we construct the Z-shape bend in Fig. 4c, and position a mirror symmetric $E_z$ source at the upper-left entrance while a detector at the exit. When the operating frequency is within the complete band gap, the electromagnetic wave will not suffer from backscattering even if it encounters the sharp corner due to the suppression of inter-valley scattering. As a result, broadband robust transport can be achieved. This is verified in Fig. 4d that the transmittance of the Z-sharp bend (red) is almost overlapped to that of straight channel (black) from 0.244-0.272 *c/a*, with a wide bandwidth of 10%. The electromagnetic field patterns at frequencies of 0.25 and 0.26 *c/a* are also plotted in Figs. 4e and 4f, so as to see the smooth round-turn passing at the two corners. Note that the transmission of the Z-shape bend drops a little near 0.24 *c*/*a* compared to that of the straight channel, as there are projected bulk states near K and K' valleys.

In conclusion, we show that valley degree of freedom can be explored to realize valley-contrasting OAM and nontrivial topology inside the all-dielectric photonic valley crystals. When the inversion symmetry is broken, the valley chiral states with intrinsic circular polarized OAM are unveiled. Selective excitation of valley chiral states is demonstrated by sources carrying OAM with proper chirality, showing valley is indeed a new degree of freedom for light manipulation. Nontrivial topology characterized with nonzero valley Chern number and topological phase transitions are illustrated. Valley dependent edge states and the associated broadband robust transport are found at the zigzag boundary of two topologically distinct photonic valley crystal. Such valley-protected transport of light are superior to the resonance-type configuration, e.g. the standard 'W1' waveguide[33,34] and see more comparison in Supplementary Information SD. Note also that the valley-protected behaviors such as valley chiral states, nontrivial topology, and broadband robust transmission, should be found in transverse-electric states with nonzero in-plane of electric fields, or other kinds of classical waves, such as acoustic wave[35]. Our work may open up a new route towards the discovery of novel fundamental states and the manipulation of spin-orbit interaction of light in nanophotonics.

This work is supported by the Natural Science Foundation of China (11274396, 11522437), the Guangdong Natural Science Funds for Distinguished Young Scholar (S2013050015694), the Guangdong special support program.

**Figures and Figure Captions**

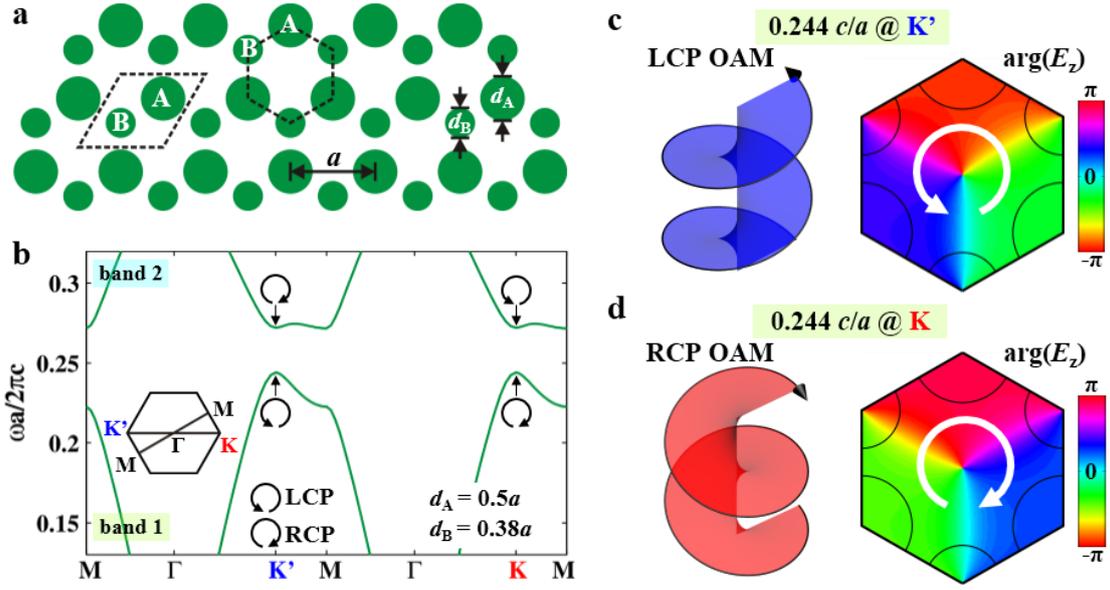

**Figure 1. Valley-contrasting orbital angular momentum (OAM) in photonic valley crystal. a**, Schematic of the inversion symmetry breaking structure. The unit-cell consists of two silicon rods embedded in air background, e.g., A and B rods in the dash-rhomboid or dash-hexagon. Both rods have the relative permittivity of $\varepsilon = 11.7$. The A rod has a diameter of $d_A = 0.5a$ while the B rod has a diameter of $d_B = 0.38a$, where $a$ is the lattice constant. **b**, Two lowest bulk bands of transverse-magnetic states. Four valley states carrying intrinsic circular polarized OAM at K' or K point are marked by counterclockwise or clockwise arrows with the corresponding chirality. Here, the left-handed (right-handed) circular polarized chirality is denoted by LCP (RCP) and the Brillouin zone with high symmetry $k$-points is shown in inset. **c-d**, Schematic of valley-contrasting chiral OAM and the phase distribution of $E_z$, i.e., $\arg(E_z)$ at the frequency of 0.244 $c/a$ at (**c**) K' and (**d**) K points. The LCP OAM state, has a counterclockwise decrease of $\arg(E_z)$, alternatively meaning that the $E_z$ field rotates counterclockwise as time evolves [see animation in Supplementary Information]. While for the RCP OAM state, it decreases clockwise reversely.

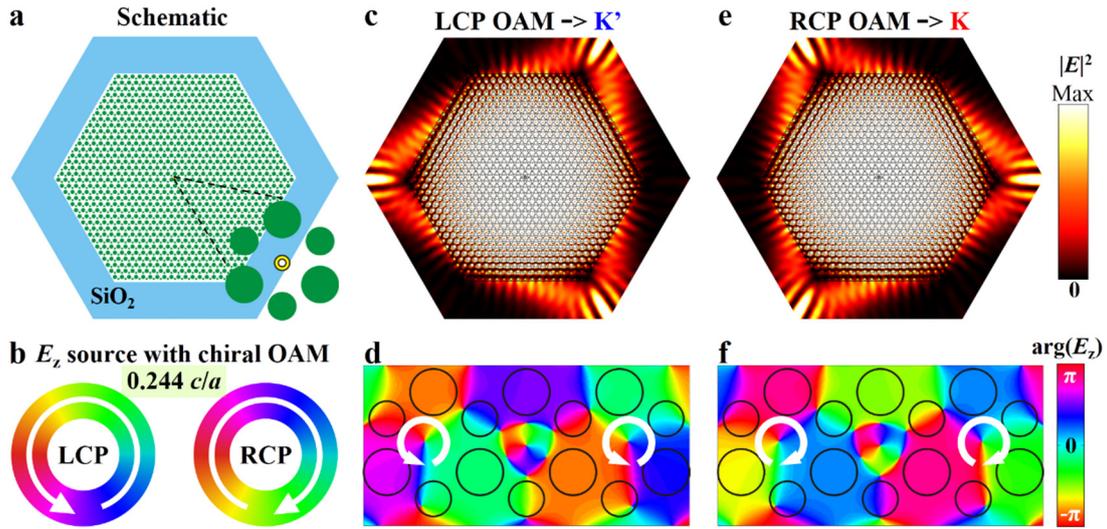

**Figure 2. Selective excitation of valley chiral states by $E_z$ source with chiral OAM.** **a**, Schematic of a hexagonal photonic valley crystal with six zigzag boundaries, surrounded by homogeneous $SiO_2$ medium. Inset: yellow marker indicates the circular $E_z$ line source around the center of the sample. **b**, Schematic of the circular $E_z$ line source with same magnitudes but phases carrying LCP OAM (left panel) or RCP OAM (right panel), with the operating frequency of 0.244 $c/a$. **c-d**, Electric density (**c**) and $E_z$ phases (**d**) of the excited K' valley state when applying the LCP source. **e-f**, Electric density (**e**) and $E_z$ phases (**f**) of the excited K valley state when the chirality of source is RCP. Note that it is experimentally detectable for the distinct differences of refracted beam at each boundary and the phase distributions inside the sample, verifying the selective excitation of valley chiral states.

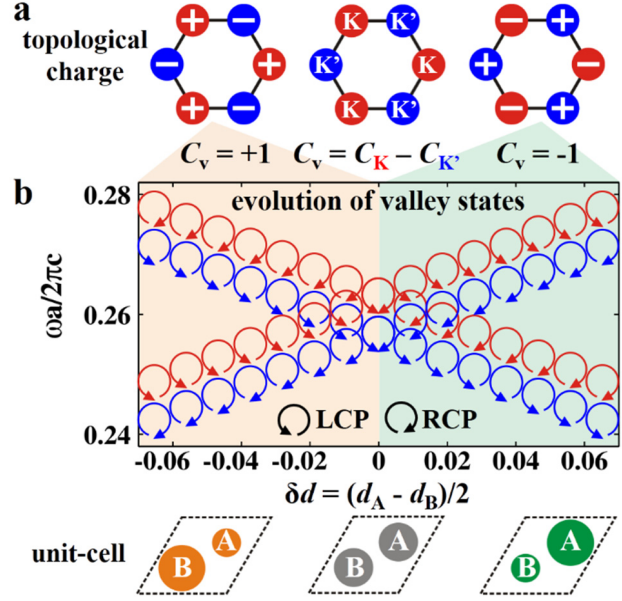

**Figure 3. Valley dependent topological charge distribution and phase transition.**
**a**, Schematics of topological charge distributions for photonic valley crystal with $\delta d < 0a$ and $\delta d > 0a$. Here, the white '+' and '-' symbols represent the topological charge of +1/2 and -1/2 in each valley, respectively. Nonzero valley Chern numbers, i.e., $C_v = (C_K - C_{K'})$ are used to distinguish the topology, and they are different for the two distinct photonic valley crystals. It indicates the topological phase transition and gives potential to the valley dependent edge states by interfacing two different photonic valley crystals. **b**, Frequency spectra of valley chiral states as a function of the amplitude of inversion asymmetry described by the term of $\delta d = (d_A - d_B)/2$. When $\delta d$, decreases, the frequencies of both LCP K' valley state and RCP K valley state boost, while those of the other two valley states drop. The state exchange occurs at $\delta d = 0a$ and results in topological phase transition. In (**b**), the degenerate frequency spectra shift a little to see the chirality of valley states, and K' and K valley chiral states are respectively shown in blue and red circular arrows.

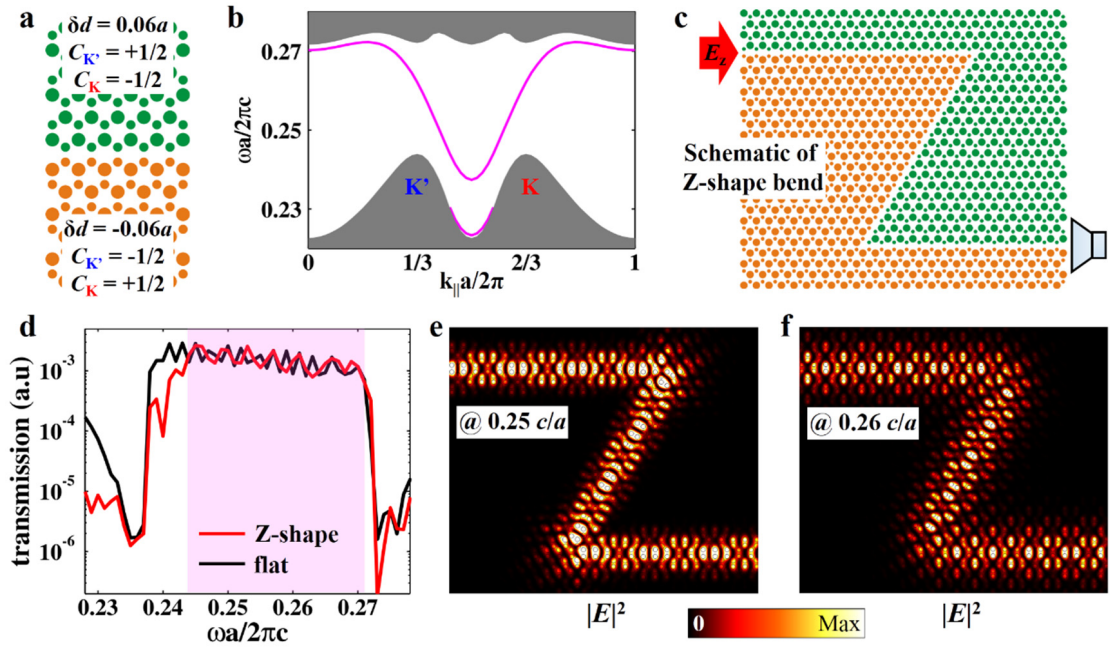

**Figure 4. Broadband robust transmission and valley dependent edge states**. **a**, Schematic of the zigzag edge constructed by photonic valley crystal with $\delta d$ = -0.06$a$ (orange) and the other with $\delta d$ = 0.06$a$ (green). **b**, Valley dependent edge states, including one with negative velocity at K' valley and the other with positive velocity at K valley. **c**, Schematic of the Z-shape bend. A mirror symmetric $E_z$ source is incident from the upper-left entrance. Transmission is measured at the lower-right exit. **d**, Broadband robust transmission from 0.244 to 0.272 $c/a$ for the comparison of the two cases of Z-shape bend (red) and straight channel (black), due to the suppression of inter-valley scattering and the vanishing field overlapping between the two valley states. **e-f,** The electric density distributions at 0.25 and 0.26 $c/a$, illustrating that the waves can pass through the sharp corners smoothly without backscattering.